\documentclass[aps,prl,twocolumn,superscriptaddress]{revtex4-1}

\usepackage{amsmath}
\usepackage{amsfonts,amssymb, graphicx}
\usepackage{wrapfig}
\usepackage{afterpage}
\usepackage{natbib}
\usepackage{color}
\usepackage[usenames,dvipsnames,svgnames,table]{xcolor}
\usepackage{amsmath}
\usepackage{amssymb}
\usepackage{graphicx}
\usepackage[usenames,dvipsnames]{xcolor}
\usepackage{tikz}
\usepackage{pgffor}
\usepackage{verbatim}
\usepackage{float}
\usepackage{mathrsfs}
\usepackage[colorlinks, breaklinks=true,linkcolor=red, citecolor=blue, linktocpage=true]{hyperref}

\newcommand{\be}{\begin{equation} }
\newcommand{\ee}{\end{equation} }
\newcommand{\ba}{\begin{eqnarray} }
\newcommand{\ea}{\end{eqnarray} }

\newcommand{\bpm}{\begin{pmatrix}}
\newcommand{\epm}{\end{pmatrix}}
\newcommand{\bmm}{\begin{matrix}}
\newcommand{\emm}{\end{matrix}}

\newcommand{\bea}{\begin{eqnarray}}
\newcommand{\eea}{\end{eqnarray}}

\newcommand{\heff}{\hbar_{\rm eff}}
\newcommand{\Kcr}{K_{\rm cr}}
\newcommand{\Ccl}{C_{\rm cl}}
\renewcommand{\Re}{{\rm Re}}

\newcommand{\llangle}{\left<\hspace{-5.83pt}\left<}
\newcommand{\rrangle}{\right>\hspace{-5.83pt}\right>}

\begin{document}

\title{Lyapunov Exponent and Out-of-Time-Ordered Correlator's Growth Rate\\ in a Chaotic System}

\author{Efim B. Rozenbaum}
\email[]{efimroz@umd.edu}
 \affiliation{Joint Quantum Institute, University of Maryland, College Park, MD 20742, USA.}
\affiliation{Condensed Matter Theory Center, Department of Physics, University of Maryland, College Park, MD 20742, USA} 
\author{Sriram Ganeshan}
 \affiliation{Simons Center of Geometry and Physics, Stony Brook, NY 11794}
\affiliation{Condensed Matter Theory Center, Department of Physics, University of Maryland, College Park, MD 20742, USA}
\author{Victor Galitski}
 \affiliation{Joint Quantum Institute, University of Maryland, College Park, MD 20742, USA.}
\affiliation{Condensed Matter Theory Center, Department of Physics, University of Maryland, College Park, MD 20742, USA} 
\affiliation{School of Physics and Astronomy, Monash University, Melbourne, Victoria 3800, Australia}

\date{\today}
\begin{abstract}
It was proposed recently that the out-of-time-ordered four-point correlator (OTOC) may serve as a useful characteristic of quantum-chaotic behavior, because in the semi-classical limit, $\hbar \to 0$, its rate of exponential growth resembles the classical Lyapunov exponent. Here, we calculate the four-point correlator, $C(t)$, for the classical and quantum kicked rotor -- a textbook driven chaotic system -- and compare its growth rate at initial times with the standard definition of the classical Lyapunov exponent. Using both quantum and classical arguments, we show that the OTOC's growth rate and the Lyapunov exponent are in general distinct quantities, corresponding to the logarithm of phase-space averaged divergence rate of classical trajectories and to the phase-space average of the logarithm, respectively. The difference appears to be more pronounced in the regime of low kicking strength $K$, where no classical chaos exists globally. In this case, the Lyapunov exponent quickly decreases as $K\to0$, while the OTOC's growth rate may decrease much slower showing higher sensitivity to small chaotic islands in the phase space. We also show that the quantum correlator as a function of time exhibits a clear singularity at the Ehrenfest time $t_E$: transitioning from a time-independent value of $t^{-1} \ln{C(t)}$ at $t < t_E$ to its monotonous decrease with time at $t>t_E$. We note that the underlying physics here is the same as in the theory of weak (dynamical) localization [Aleiner and Larkin, Phys. Rev. B {\bf 54}, 14423 (1996); Tian, Kamenev, and Larkin, Phys. Rev. Lett. {\bf 93}, 124101 (2004)] and is due to a delay in the onset of quantum interference effects, which occur sharply at a time of the order of the Ehrenfest time.
\end{abstract}

\maketitle

{\it Introduction. ---} One of the central goals in the study of quantum chaos is to establish a correspondence principle between classical and quantum dynamics of  classically chaotic systems~\cite{Casati79, Shepelyansky83, *Casati86, *Dittrich90, Berman78, *Berry79, Chirikov81, Chirikov88, berry1983semiclassical, Haake87, *Haake10}.  Several previous works~\cite{Fishman82, *Fishman82_2, *Prange84, Toda87, Haake92, Alicki96, Haake10} have attempted  to recover fingerprints of classical chaos in quantum dynamics. In particular, Aleiner and Larkin~\cite{aleiner1996divergence, *aleiner1997divergence_2, *aleiner2000shotnoise} showed the existence of a semiclassical ``quantum chaotic'' regime attributed to the delay in the onset of quantum effects (due to weak localization) revealing the key measure of classical chaos -- the Lyapunov exponent (LE). Recently, the subject of quantum chaos has been revived by the discovery of an unexpected conjecture that puts a bound on the growth rate of an out-of-time-ordered four-point correlator (OTOC)~\cite{kitaev, Maldacena16}. OTOC was first introduced by Larkin and Ovchinnikov to quantify the regime of validity of quasi-classical methods in the theory of superconductivity~\cite{Larkin69}. The growth rate of OTOC appears to be closely related to LE. Recent works have proposed experimental protocols to probe OTOC in cold atom and cavity QED setups~\cite{swingle2016measuring, *yao2016interferometric}. Several recent preprints have employed OTOC as a probe to characterize many-body-localized systems~\cite{huang2016out,*chen2016quantum,*swingle2016slow,*fan2016out}.

In this letter, we calculate the Lyapunov exponent, OTOC and the two-point correlator for the quantum kicked rotor (QKR), which is a canonical driven model of quantum chaos~\cite{Casati79, Izrailev80, *Izrailev80_2, Chirikov81}. The classical version of this model manifests regular-to-chaotic transition (as a function of driving strength $K$) which enables us to benchmark the behavior of OTOC against the presence and absence of classical chaos. We show that in the limit of small dimensionless effective Planck's constant, $\heff \to 0$, there exists a ``quantum chaotic'' regime~\cite{Larkin69, aleiner1996divergence, *aleiner1997divergence_2, *aleiner2000shotnoise} at early times where OTOC, $C(t) = - \left\langle \left[ \hat{p}(t),\, \hat{p}(0) \right]^2 \right\rangle$, grows exponentially. This correlator's growth rate, $\tilde{\lambda}$, that we abbreviate for brevity as CGR, is found to be independent of the dimensionless Planck's constant, $\heff$, and is purely classical at early times for the kicked rotor. Most importantly, the CGR and the standard definition of LE in classical systems are shown to be different at all non-zero kicking strengths. In particular, for the classically regular regime, $K < \Kcr$, CGR significantly exceeds LE due to much higher sensitivity to the presence of small chaotic islands. For the classically deeply chaotic regime, $K \gg \Kcr$, CGR exceeds LE by nearly a constant. We attribute these distinctions to different averaging procedures carried out to extract these exponents and posit that this statement may be more general than the specific QKR model studied here.

We also show that deviations from the essentially classical behavior of OTOC, $C(t)\sim e^{2\tilde{\lambda} t}$, occur sharply at a time of the order of the Ehrenfest time, $t_E$, where OTOC exhibits a clear cusp. This corresponds to the minimal time it takes classical trajectories to self-intersect indicating the onset of quantum interference effects~\cite{aleiner1996divergence, *aleiner1997divergence_2, *aleiner2000shotnoise}. This is in analogy to the weak dynamical localization discussed by Tian et al.~\cite{tian2004weak}. At longer times, $t  > t_E$, the quantum disordering effects subdue the exponential growth dictated by the CGR to a power-law growth. 

Finally, we calculate the two-point correlation function and show that CGR, $\tilde{\lambda}$, is not revealed in this quantity (nor in the single-point average -- e.g., the kinetic energy as has been well known~\cite{Haake10}). However, we find that the two-point correlator does contain fingerprints of classical transition from regular dynamics to chaos even deep in the quantum regime at long times, which has been a subject of long-standing theoretical and experimental interest~\cite{Hensinger01,Steck01,Lemos12,Larson13}.

{\it Quantum Kicked Rotor. ---} The dimensionless Hamiltonian of QKR \cite{Casati79, Izrailev80, Izrailev80_2, Chirikov81} can be written as
\begin{align}
\hat{H} = \dfrac{\hat p^2}{2} + K\cos(\hat x)\Delta(t),
 \label{eq:Hamiltonian}
\end{align}
where $\Delta(t) = \sum\limits_{j = -\infty}^{\infty} \delta(t - j)$ is the sum of $\delta-$pulses, $\hat p$ is the dimensionless angular-momentum operator, $\hat x$ is the angular coordinate operator, and $t$ is the dimensionless time. The QKR is characterized by two parameters. One of them, the kicking strength $K$, comes from the classical kicked rotor (KR, also called Chirikov standard map)~\cite{Chirikov79}. Another parameter is the dimensionless effective Planck constant $\heff$, which enters the dimensionless angular momentum operator ($\hat{p} = -i\heff\frac{\partial}{\partial x}$) and the dimensionless Schr\"{o}dinger equation: $i\heff\frac{\partial}{\partial t}\left| \Psi \right> = \hat{H}\left| \Psi \right>$. The eigenvalues of $\hat{p}$ are quantized in units of $\heff$ due to the periodic boundary conditions. Note that in the classical KR, the parameter $\heff$ is absent. In order to understand how classical chaos emerges from quantum dynamics, we compute OTOC and the two-point correlator in the regime of $\heff\rightarrow 0$ at short time scales. 

{\it Lyapunov Exponent and OTOC's growth rate (CGR). ---}  To specify our quantum diagnostics for chaotic behavior in the QKR, we choose OTOC, $C(t)$
\cite{Larkin69, Maldacena16}, and two-point correlator, $B(t)$, as:
\begin{align}
C(t) = -\left\langle \left[ \hat{p}(t), \hat{p}(0) \right]^2 \right\rangle, \quad  B(t) = \Re\left\langle \hat p(t) \hat p(0)\right\rangle.
\label{fourandtwopt}
\end{align}
We point out that $C(t)$ is closely related to the Loschmidt echo (also known as fidelity). In the previous works, fidelity has been used as a theoretical and experimental diagnostic of quantum chaos~\cite{Peres84,Pastawski00,Jacquod01,Jalabert01,
Cucchietti02,Cucchietti02echo,Chaudhury09,Garcia_Mata11,swingle2016measuring}. 

\begin{figure} 
\includegraphics[width=\linewidth]{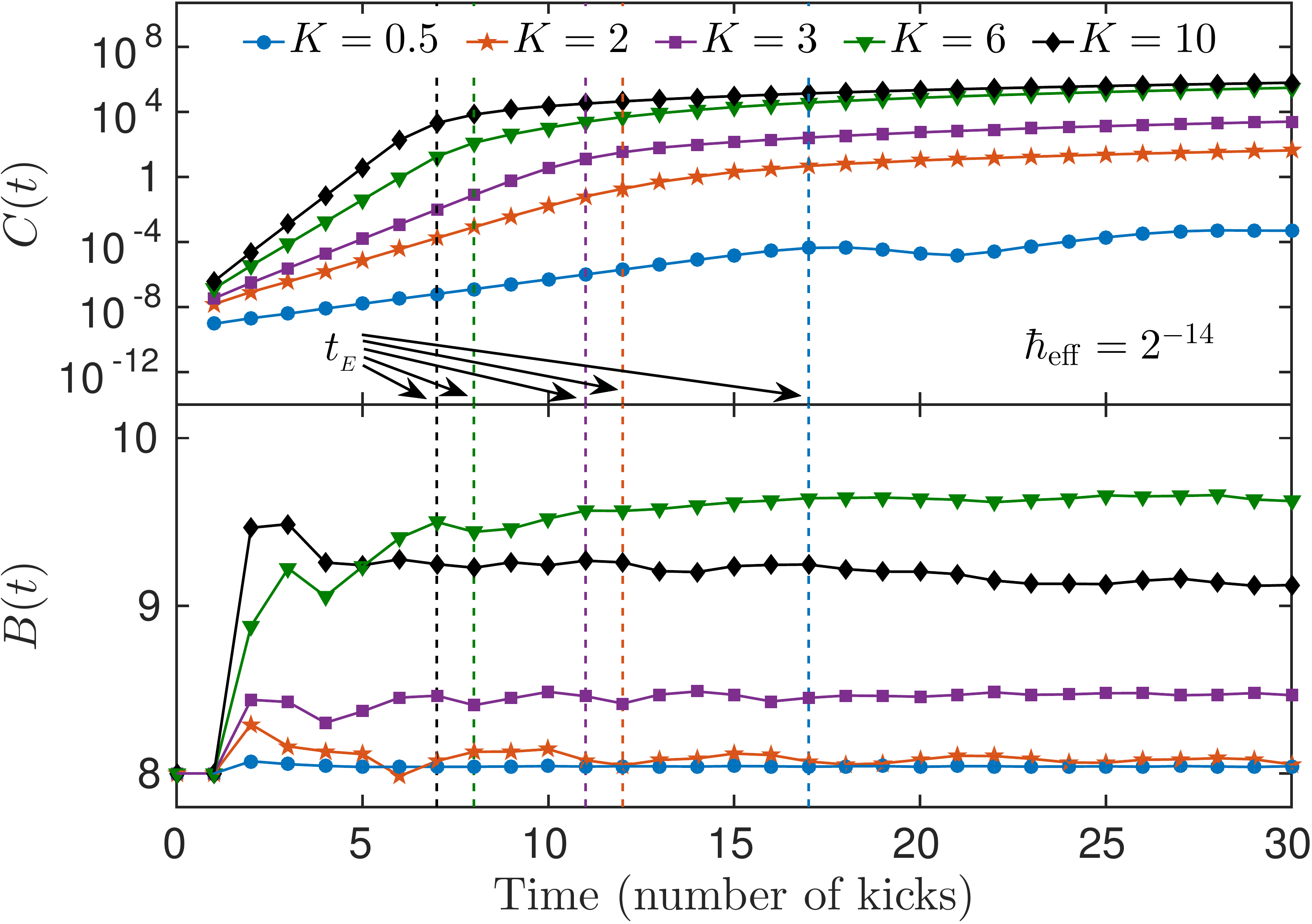}
\caption{(Color online) The upper panel shows OTOC, $C(t)$, vs $t$ in the semi-log scale for various values of the kicking strength ($K=0.5, 2, 3, 6, 10$) and $\heff=2^{-14}$. The lower panel is a plot of the two-point function, $B(t)$, vs $t$ at the corresponding parameters (in the linear scale). Averaging is performed over the Gaussian wave packet defined in Eq.~(\ref{eq:Gauss_wf}) with $p_0 = 0$ and $\sigma=4$.}\label{4ptV2pt} 
\end{figure}
Before carrying out quantum calculations, we consider the classical correspondence of  $C(t)$~\cite{Larkin69,Maldacena16}. At short times $t < t_E$ \cite{SuppCvsCcl}:
\begin{equation}\label{eq:Mald_pppp_quant_to_class}
\hspace{-5pt} C(t) = \heff^2\left<\left(\dfrac{\partial\hat{p}(t)}{\partial x(0)}\right)^2 \right> \approx \heff^2\llangle\left(\dfrac{\Delta p(t)}{\Delta x(0)}\right)^2 \rrangle = \Ccl(t),
\end{equation}
where we changed the expectation value of the operator derivative to the finite differences of the classical variables averaged over the phase space ($\,\,\llangle \,\,\ldots\,\, \rrangle$ denotes classical phase-space average). Note that the averaging allows for direct comparison of the classical $\Ccl(t)$ to the quantum $C(t)$. Such a comparison would not always be possible for local quantities because of quantum wave-packet spreading. Due to the presence of chaotic regions in the phase space, $\Ccl(t) \sim e^{2\tilde{\lambda} t}$ grows exponentially. Now we compare this classical CGR, $\tilde{\lambda} = \lim\limits_{t \to \infty}\lim\limits_{\Delta x(0) \to 0} \frac{1}{2t}\ln\frac{\Ccl(t+1)}{\Ccl(1)}$, to the standard definition of the LE: $\lambda = \llangle\lim\limits_{t \to \infty}\lim\limits_{d(0) \to 0} \frac{1}{t}\ln\frac{d(t)}{d(0)}\rrangle$ \cite{SuppLyap} (where $d(t) = \sqrt{[\Delta x(t)]^2 + [\Delta p(t)]^2}$). Notice that there are key differences between definitions of $\lambda$ and $\tilde{\lambda}$ coming from the different orders of squaring, averaging, taking ratio and applying logarithm.

Next, we proceed to check if the classical correspondence follows through in a quantum calculation of $C(t)$ and compare the rate of exponential growth of $C(t)$  to $\tilde{\lambda}$ extracted from $\Ccl(t)$ and to LE $\lambda$. For the quantum case, the averaging in Eq.~(\ref{fourandtwopt}) is performed in the Schr\"odinger picture over some initial state $\left| \Psi(0) \right>$. We use individual angular-momentum eigenstates $\left| \Psi(0) \right> = \left|n\right>: \; \hat{p}\left|n\right> = \heff n\left|n\right>$ and Gaussian wave-packets:
\begin{equation}\label{eq:Gauss_wf}
\hspace{-10.3pt}\left|\Psi(0)\right> = \sum\limits_{n=-\infty}^{\infty}\hspace{-5pt}a^{(0)}_n\left|n\right>,\;\; a^{(0)}_n \sim \exp\left[-\dfrac{\heff^2(n-n_0)^2}{2\sigma^2}\right],
\end{equation}
where $n_0 = p_0/\heff$. In this calculation, we use wave-packet (\ref{eq:Gauss_wf}) with $p_0 = 0$ and $\sigma = 4$. Numerically, $\left|\Psi\right>$ is represented in a finite basis of eigenstates $\left|n\right>$, $n\in[-N; N-1]$. All functions of only $\hat{p}$ are applied in this basis, and all functions of only $\hat{x}$ are applied in the Fourier-transformed representation. We use adaptive grid with $2\heff N \in [2^7; 2^{16}]$, so that all physical observables are well converged. The wave-function is evolved by switching between representations back and forth and applying the Floquet operator $\hat{F} = e^{-i\hat{p}^2/2\heff}e^{-iK\cos(\hat{x})/\heff}$ in parts. Then the correlators are calculated in the Schr\"{o}dinger picture.
 \begin{figure} 
\includegraphics[width=\linewidth]{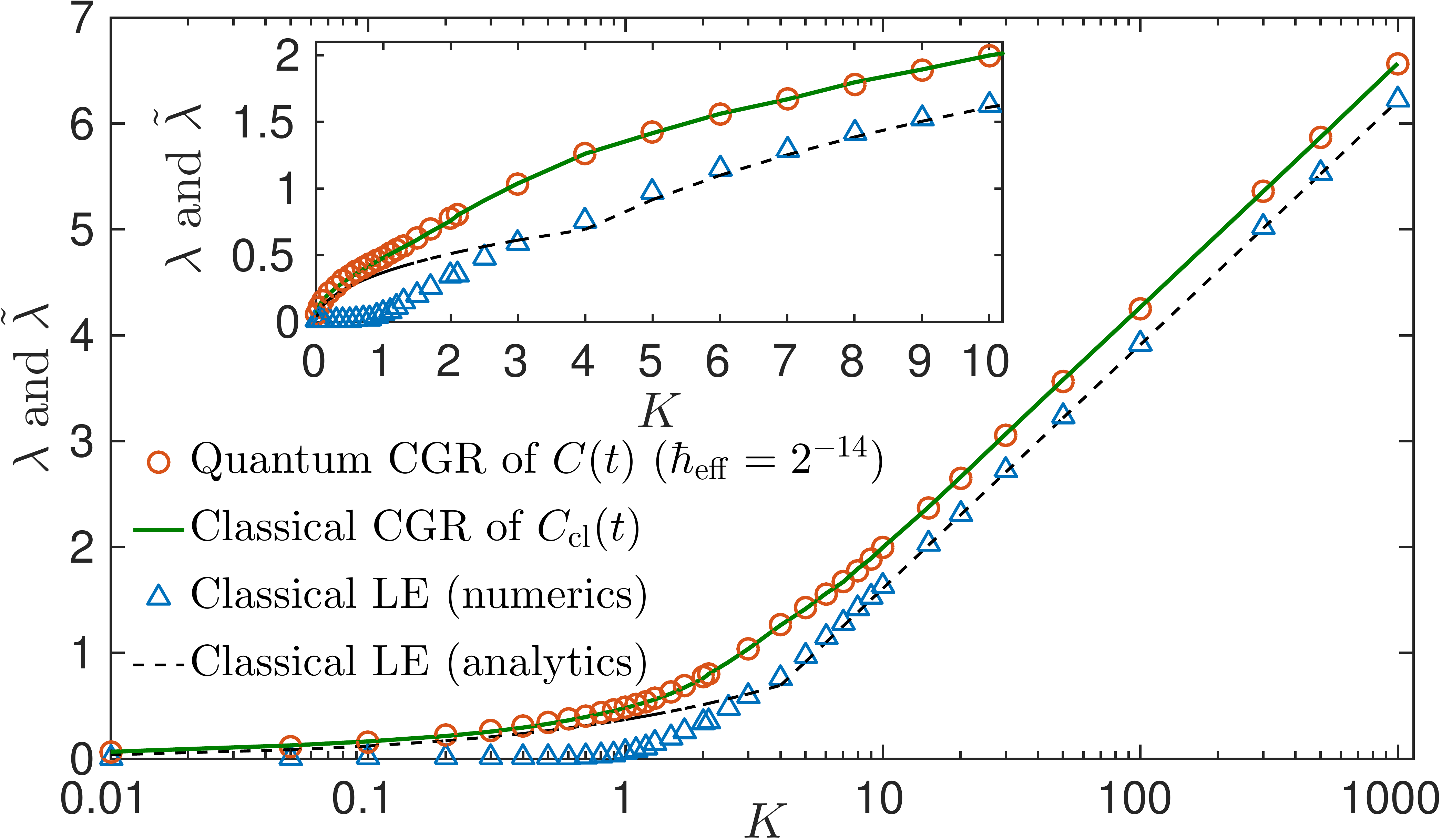}
\caption{(Color online) Red circles: early-time growth rate of $C(t)$ at $\heff = 2^{-14}$ (quantum CGR). The rest of the data is classical. Green solid line: growth rate of $\Ccl(t)$ (classical CGR). Blue triangles: LE calculated numerically. Black dashed line: LE according to the Chirikov's analytical formula (\ref{eq:Lyapunov_analytic}). The main plot and the inset show the same data in lin-log and linear scales, respectively (and in different ranges). At $K \gtrsim 8$, the difference between CGR and LE is constant $\approx\ln\sqrt{2}$. The initial state in $C(t)$ is the Gaussian (\ref{eq:Gauss_wf}) with $p_0 = 0$ and $\sigma=4$. Fitting details for extracting CGR from $C(t)$ and $\Ccl(t)$ are given in the main text.}
\label{LE_vs_CGR} 
\end{figure}

The exponential growth of $C(t)$ lasts between the time $t_d$ and the Ehrenfest time $t_E$~\cite{Berman78, *Berry79, Maldacena16}. To achieve a hierarchical separation between $t_d$ and $t_E$ ($\frac{t_E}{t_d}\gg 1$) for the QKR, we have to tune both $K$ and $\heff$. The  estimates of $t_d\sim [\ln (K/2)]^{-1}$ and $t_E\sim \frac{|\ln \heff|}{\ln(K/2)}$ at $K>4$ guide our choice of parameters to achieve this separation.  The smallest $\heff$ within the scope of our numerics is $\heff=2^{-14}$. For this value of $\heff$, the Ehrenfest time is in the range $7\lesssim t_E \lesssim 17$ kicks for the range of kicking strength $0.5\leq K\leq10$. By $K = 1000$, $t_E$ shrinks down to $3$ kicks, but at these values of $K$, it appears to be enough to extract a well averaged exponent. For the above mentioned parameter regimes, we numerically observe the exponential growth of $C(t)$ at early times ($t<t_E$) as shown in the Fig.~\ref{4ptV2pt}, upper panel. Fig.~\ref{4ptV2pt} also shows that $t_E$ decreases upon increasing the kicking strength $K$ for fixed $\heff$.  In contrast to $C(t)$, the two-point correlator $B(t)$ saturates at time $t\sim 2$ kicks (Fig.~\ref{4ptV2pt}, lower panel). 

Equipped with the early time behavior of $C(t)$, we are in a position to extract the rate of its exponential growth, i.e obtain CGR from the quantum calculation. We carry out a four-pronged comparison between CGR from the quantum calculation of $C(t)$, CGR from the classical calculation of $\Ccl(t)$, numerically obtained LE for KR and analytical estimates (\ref{eq:Lyapunov_analytic}) of LE from Chirikov's standard map analysis \cite{Chirikov79}. The Chirikov's analytical formula reads: \vspace{-10pt}
\begin{equation}\label{eq:Lyapunov_analytic}
\lambda \approx \dfrac{1}{2\pi}\int\limits_{-\pi}^\pi dx \; \ln{L(x)},
\end{equation}
where 
\begin{equation}\label{eq:Lyapunov_analytic_continued}
L(x) = \left|   1 + \dfrac{k(x)}{2} + {\rm sgn}[k(x)]\sqrt{k(x)\left(1+\dfrac{k(x)}{4}\right)  }   \right|
\end{equation}
and $k(x) = K\cos x$. The simplified expression $\lambda \approx \ln(K/2)$ valid at large $K$ is obtained by substituting $L(x) \approx |k(x)|$ into Eq.~(\ref{eq:Lyapunov_analytic}) \cite{Chirikov79, SuppLyap}.

In Fig.~\ref{LE_vs_CGR}, we compare the exponents  obtained in four ways listed above. In order to extract the exponents from $C(t)$, we determine the times, after which the exponential growth starts slowing down, and fit $C(t)$ from $t=1$ up to these times to the function $a e^{2\lambda_{\rm fit} (t-1)}$ to find the parameter $\lambda_{\rm fit}$ ($C(0) = 0$, so we omit $t=0$).
\begin{figure} 
\includegraphics[width=\linewidth ]{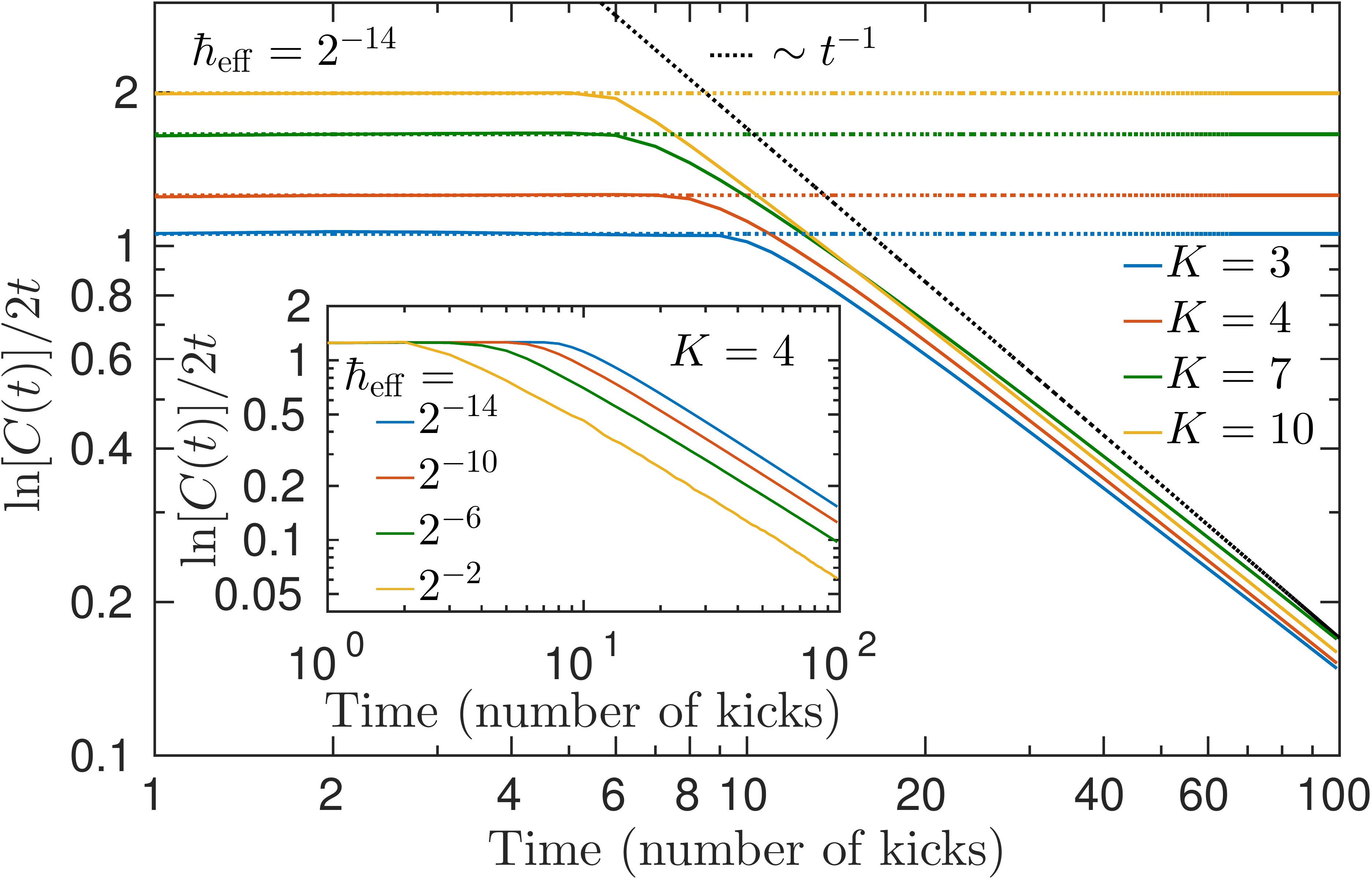}
\caption{(Color online) Main plot: $\ln[C(t)]/2t$ vs $t$ in the log-log scale for $K=3,4,7,10$ (from bottom to top line, respectively) and $\heff = 2^{-14}$. The flat region at early times quantifies the exponential growth rate of $C(t)$. This flat region persists up to the time $t_E$, at which the exponential growth slows down to a power-law growth with a slowly decreasing power. Dotted lines are the eye guides: horizontal lines extend the flat regions, sloped line is shown for power comparison. Inset: $\ln[C(t)]/2t$ vs $t$ in the log-log scale for $K=4$ and $\heff = 2^{-14}, 2^{-10}, 2^{-6}, 2^{-2}$ (from top to bottom line, respectively). The rate of exponential growth is the same for different values of $\heff$, but $t_E$ shrinks when $\heff$ increases.}
\label{Lyapunov_t} 
\end{figure}
Numerical calculations of the classical LE and of the classical CGR [i.e. the growth rate of $\Ccl(t)$] are performed using the map tangent to the standard map -- this standard procedure is outlined in the supplemental material \cite{SuppLyap}. Notice that the exponents extracted from $C(t)$ (quantum CGR) and from $\Ccl(t)$ (classical CGR) are in an excellent agreement for all values of $K$. Both classical and quantum CGRs significantly exceed LE at $K < \Kcr$. This indicates that CGR may not be a reliable tool for discriminating between classically regular and chaotic dynamics in a global sense, but it can be employed to detect the existence of local disconnected chaotic islands more efficiently than LE. As expected, numerically calculated values and analytical estimates of the classical LE agree with each other for $K \gtrsim 3$.  At large $K$, the difference between CGR and LE becomes nearly constant $\approx \ln\sqrt{2}$. We attribute this distinction primarily to the difference in the order of averaging in CGR and LE. 

\begin{figure}
\includegraphics[width=\linewidth]{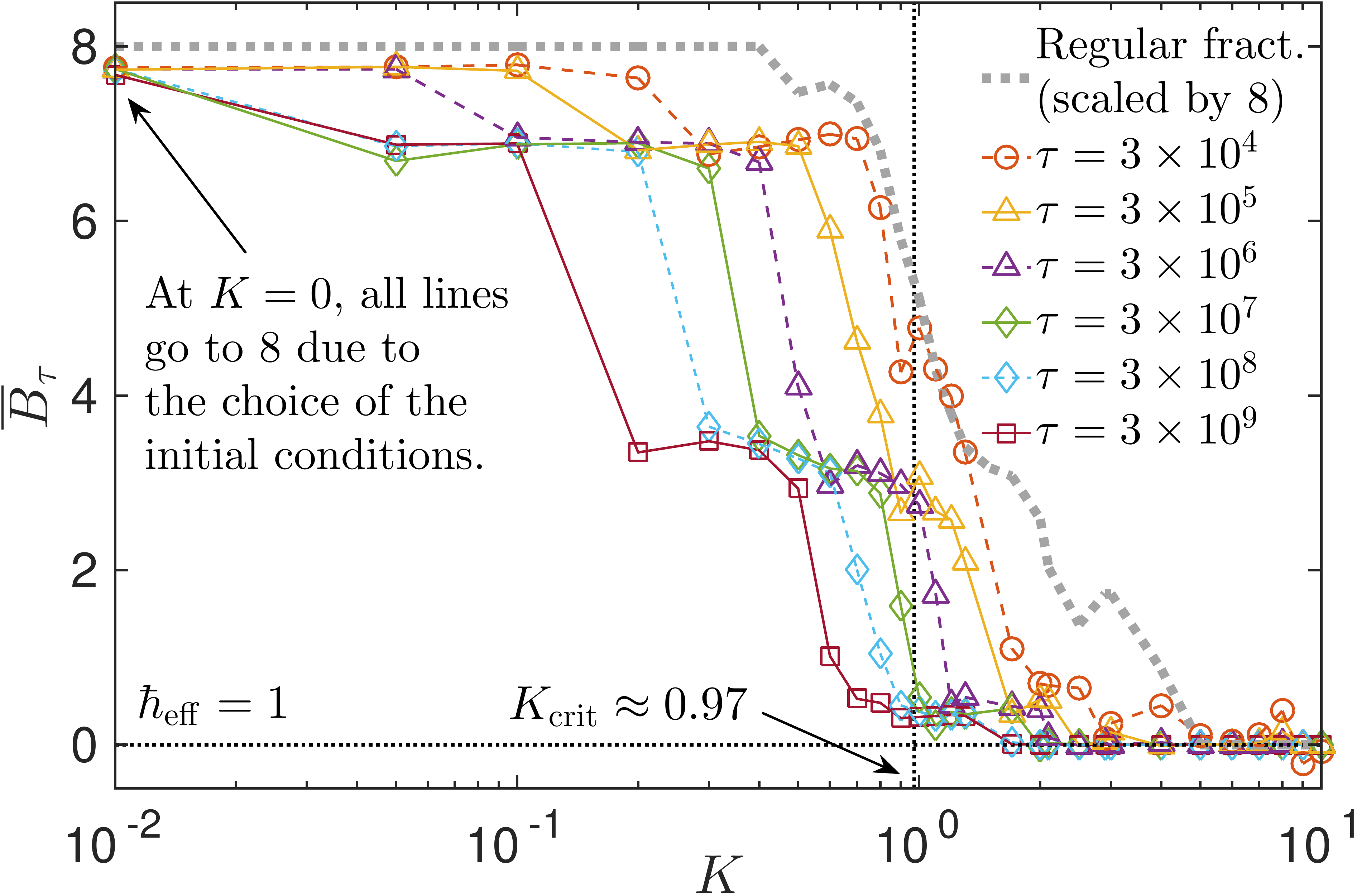}
\caption{\label{fig:pp_and_reg_cont} (Color online) Long-time average $\overline{B}_\tau$ (\ref{twopoint}) (over various windows $\tau$) of the two-point correlator $B(t)$ as a function of $K$ compared to the regular fraction of the phase space weighted with the initial Wigner distribution $P(x,p)$ (scaled). The trend with increasing $\tau$ shows that at all $K \neq 0$, the correlations decay in time, but the rate of this decay has a step-like dependence on $K$. At $K > \Kcr$, the decay is quite fast, while at $K < \Kcr$, it takes $\overline{B}_\tau$ at least exponentially large window to vanish. It is not clear from the data whether at small $K\neq 0$, averaged correlator eventually goes to zero at $\tau \to \infty$ or is bounded from below. Initial state corresponding to $P(x,p)$ is the Gaussian (\ref{eq:Gauss_wf}) with $p_0 = 0$ and $\sigma=4$.}
\end{figure}
Now we proceed to consider the deviation of $C(t)$ from its classical counterpart $\Ccl(t)$ that manifests sharply at a time close to $t_E$. The onset of this deviation in OTOC is closely related to the weak dynamical localization effects~\cite{tian2004weak}. In Fig.~\ref{Lyapunov_t}, we plot $\ln [C(t)]/2t$ as a function of time $t$ in the log-log scale. This plot is constant [corresponding to the exponential rise of $C(t)$] at early times. Beyond $t_E$, the exponential growth slows down to a power-law growth (nearly quadratic growth around $t\sim 100$ kicks). At long times, the growth of $C(t)$ slows down further, but numerics quantifying this slowdown is out of the scope of the present manuscript. However, we can unambiguously extract the exponent associated with the exponential growth prior to $t_E$. Note that in the range of $K$ and $\heff$ where the region of the exponential growth of $C(t)$ is present ($t_E \geq 3$), $\tilde{\lambda}$ does not depend on $\heff$ (see Fig.~\ref{Lyapunov_t}, inset). 
\begin{figure}
\includegraphics[width=\linewidth]{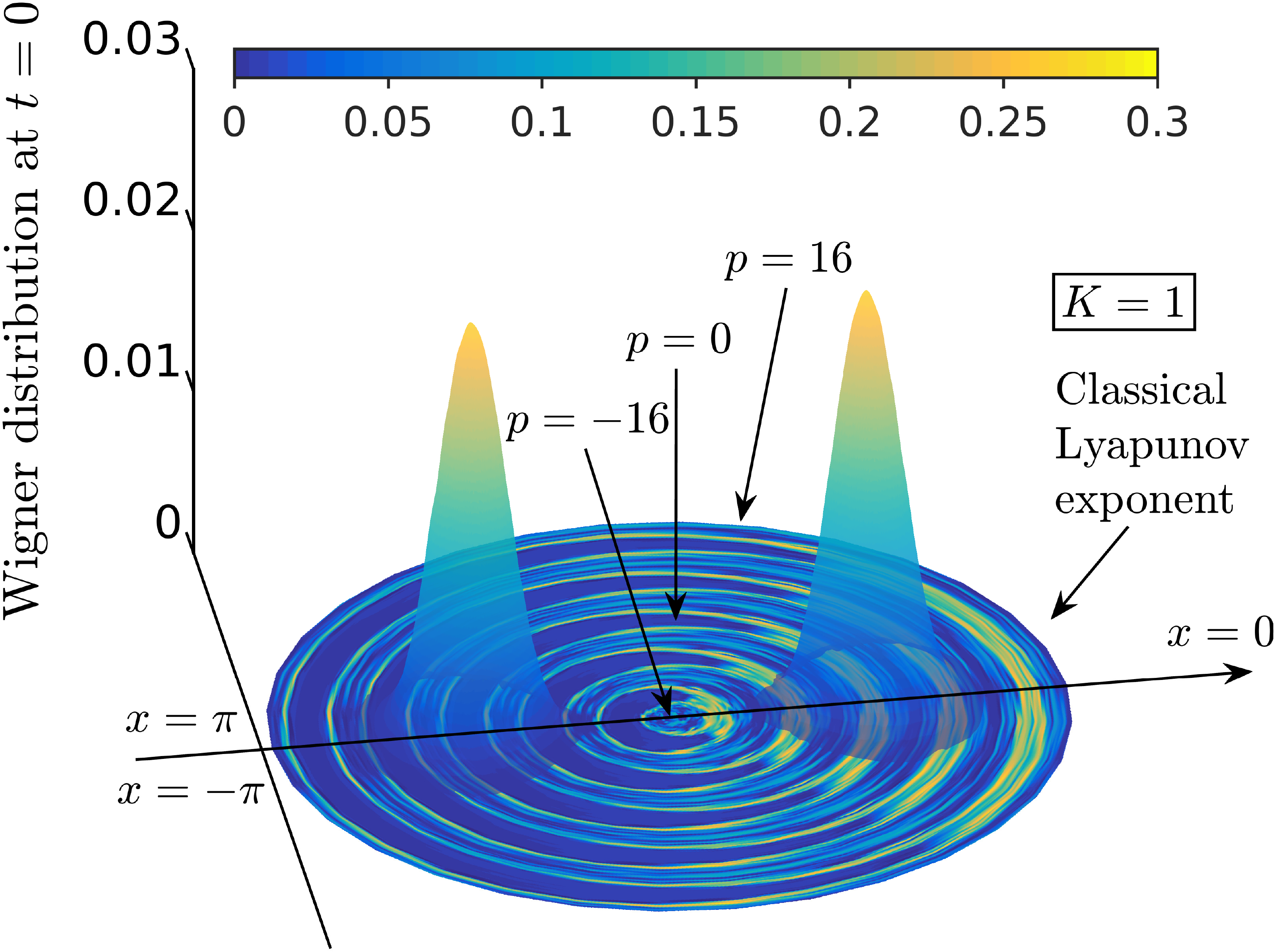}
\caption{\label{fig:Wigner_vs_Phase_Space} (Color online) Initial Wigner distribution $P(x,p)$ (3D plot) on the top of the classical Lyapunov exponent (shown in color in the horizontal plane, see colorbar for numerical values). Initial state corresponding to $P(x,p)$ is the Gaussian (\ref{eq:Gauss_wf}) with $p_0 = 0$ and $\sigma=4$. Lyapunov exponent is shown for $K = 1$.}
\end{figure}

{\it Regular-to-chaotic transition in long-time quantum dynamics. ---} Classical KR is famous for its transition from regular motion to chaotic behavior that occurs as $K$ is increased above $K = \Kcr \approx 0.97$. The chaotic phase is characterized by the quasi-random walk in the angular-momentum space that leads to diffusion in angular momentum, so that the rotor's energy averaged over the phase space grows linearly with time (number of kicks). On the other hand, at long times QKR undergoes dynamical localization (which is closely connected to Anderson localization in disordered solids~\cite{Fishman82}) and around $\heff \sim 1$, the standard diagnostic -- the average energy, i.e. the one-point correlator -- seems insensitive to the presence or absence of classical chaos~\cite{Casati79, Chirikov81}. Thus a question arises: is there a quantum diagnostic that manifests a robust signature of regular-to-chaotic classical transition in the purely quantum dynamics even in the dynamically localized regime ($\heff=1$, $t_d \gg t_E$)? Remarkably, the two-point correlator [$B(t)$ in Eq.~(\ref{fourandtwopt})] contains a sharp signature of the classical transition \footnote{In addition, time-dependence of $B(t)$ in the interval $t \in [0, t_{\rm max}]$ is much more accessible than that of OTOC $C(t)$, as their computation complexities scale as $O(t_{\rm max})$ and $O(t_{\rm max}^2)$, respectively.}. In particular, we consider $B(t)$ averaged over time within various windows $\tau$:
\begin{align}
 \overline{B}_\tau=\frac{1}{\tau}\sum\limits_{t=0}^\tau\Re\left<p(t)p(0)\right>.
\label{twopoint}
\end{align}
As shown in Fig.~\ref{fig:pp_and_reg_cont}, this averaged correlator maintains a sharp step-like structure as a function of $K$ for several orders of magnitude in $\tau$ (we reached as large window as $\tau = 3\times 10^9$, which is many orders of magnitude longer than any characteristic time scale in the system). This implies that at very long times, the quantum system does not loose the information about the classical transition. The plot supports the following very intuitive statement. The larger the chaotic fraction of the classical phase space is, the shorter the correlation-decay time window becomes (for explicit demonstration of this behavior, the dependence of $\overline{B}_\tau$ on the averaging window size $\tau$ is given in Fig.~S2 in the supplementary material). Therefore, we can relate $\overline{B}_\tau$ to the regular part of the phase space weighted by the initial Wigner distribution $P(x,p)$ of QKR (see Fig.~\ref{fig:Wigner_vs_Phase_Space} for illustration). However, $\overline{B}_\tau$ keeps decaying with time, while the regular phase-space fraction is a constant determined by the initial conditions and $K$, so a fixed window should be chosen for comparison. As the ratio of regular to chaotic areas of the phase space decreases, so does the average value of the correlator over a this window, until it reaches zero at large $K$, where almost no regular regions are present.

\begin{acknowledgments}
{\it Acknowledgments. ---} ER and VG were supported by NSF-DMR 1613029 and the Simons Foundation. SG gratefully acknowledges support by LPS-MPO-CMTC, Microsoft Station Q. ER and VG are grateful to Shmuel Fishman for discussions and valuable comments. SG acknowledges valuable discussions with A. G. Abanov and D. Huse. VG is grateful to Chushun Tian and Hui Zhai for valuable discussions and hospitality at the Institute for Advanced Study, Tsinghua University, where a part of this work was completed.
\end{acknowledgments}

\bibliography{LE_CGR_in_QKR}

\clearpage

\pagebreak
\section{Supplemental Material}

\setcounter{equation}{0}
\setcounter{figure}{0}
\setcounter{table}{0}
\setcounter{page}{1}
\makeatletter
\renewcommand{\theequation}{S\arabic{equation}}
\renewcommand{\thefigure}{S\arabic{figure}}

{\it Classical Lyapunov exponent. ---} Classical chaotic systems, such as KR, are characterized by Lyapunov exponents that determine the rate of exponential separation of initially close trajectories in the phase space. In one-dimensional systems, there is only one positive Lyapunov exponent, $\lambda$, given by the phase-space average of $\lambda(x,p)$:
\begin{equation}
d(t) \approx d(0)e^{\lambda(x,p)t},
\end{equation}
where $d(t) = \sqrt{\left[x'(t) - x(t)\right]^2 + \left[p'(t) - p(t)\right]^2}$ is the distance at time $t$ between two initially close trajectories in the phase space. This can be used to extract $\lambda$ as:
\begin{equation}\label{eq:lambda_conventional}
\lambda = \left<\hspace{-2.5pt}\left< \lambda(x,p) \right>\hspace{-2.5pt}\right> = \llangle \lim\limits_{t \rightarrow \infty}\lim\limits_{d(0) \rightarrow 0} \dfrac{1}{t}\ln\dfrac{d(t)}{d(0)} \rrangle.
\end{equation}

{\it Chirikov's analytical derivation of formula} (5) {\it for Lyapunov exponent. ---} Consider two trajectories that obey the standard map:
\begin{eqnarray}\label{eq:standard_map}
\vspace{-7pt}
\begin{cases}
p_{n+1} = p_n + K\sin x_n\\
x_{n+1} \hspace{-7pt}\underset{\hspace{-4pt}\mod 2\pi}{=}\hspace{-7pt} x_n + p_{n+1}
\end{cases}, \\
\begin{cases}
p'_{n+1} = p'_n + K\sin x'_n\\
x'_{n+1} \hspace{-7pt}\underset{\hspace{-4pt}\mod 2\pi}{=}\hspace{-7pt} x'_n + p'_{n+1}
\end{cases}.\vspace{10pt}
\end{eqnarray}
Let us introduce relative coordinates: $\xi_n = x'_n-x_n$ and $\eta_n = p'_n - p_n$, so that ${\boldsymbol d}(n) = \dbinom{\eta_n}{\xi_n}$. Standard map results for them in:
\begin{equation}\label{eq:standard_map_relative}
\begin{cases}
\eta_{n+1} = \eta_n + K (\sin x'_n - \sin x_n)\\
\xi_{n+1} = \xi_n + \eta_{n+1}
\end{cases},
\end{equation}
where $x_n$ and $x'_n$ cannot be eliminated exactly. Using a trigonometric identity, we can rewrite:
\begin{eqnarray}\label{eq:sin_transform}\nonumber
&&\sin x'_n - \sin x_n = \sin (x_n + \xi_n) - \sin x_n \\
&&= \sin x_n (\cos \xi_n - 1) + \sin \xi_n \cos x_n.
\end{eqnarray}
Consider a mapping tangent to that in Eq.~(\ref{eq:standard_map_relative}). For that, assume $\xi_n$ is small. Then, to the linear order in $\xi_n$, the expression in Eq. (\ref{eq:sin_transform}) is equal to $\xi_n\cos x_n$, so that the tangent mapping is:
\begin{equation}\label{eq:standard_map_relative_tangent}
\begin{cases}
\eta_{n+1} = \eta_n + (K \cos x_n)\xi_n \\
\xi_{n+1} = \xi_n + \eta_{n+1}
\end{cases}.
\end{equation}
It still contains $x_n$ determined by the standard map and thus non-linearly dependent on time, but there is a class of trajectories for which this mapping is linear: periodic trajectories with $x_n \equiv 0$ or $\pi$ and $p_n \in 2\pi\mathbb{Z}$. For them, the standard mapping is trivial: $p_{n+1} = p_n,\; x_{n+1} = x_n$, and $k \equiv K\cos x_n \equiv \pm K$ (for $x_n \equiv 0$ or $\pi$, respectively). Consider the mapping (\ref{eq:standard_map_relative_tangent}) for trajectories near these special ones and rewrite it in the matrix form:
\begin{equation}\label{eq:standard_map_relative_tangent_linear}
\dbinom{\eta_{n+1}}{\xi_{n+1}} = 
\begin{pmatrix}
1 & k \\ 1 & k+1
\end{pmatrix}
\dbinom{\eta_{n}}{\xi_{n}}.
\end{equation}
The length of ${\boldsymbol d_n} = \dbinom{\eta_{n}}{\xi_{n}}$ is the distance between two trajectories in the phase space at step $n$. Denote eigenvalues of the matrix in Eq.~(\ref{eq:standard_map_relative_tangent_linear}) as $\ell_{\pm}$ and the corresponding eigenvectors as ${\boldsymbol{\it e_{\pm}}}$. Let us expand ${\boldsymbol d_n} = u_n {\boldsymbol{\it e_+}} + v_n{\boldsymbol{\it e_-}}$. Then for the coefficients, we have: $u_{n+1} = \ell_+u_n, \; v_{n+1} = \ell_-v_n$, so that $u_n = \ell_+^nu_0, \;v_n = \ell_-^nv_0$. Eigenvalues $\ell_\pm$ are:
\begin{equation}
\ell_\pm = 1 + \frac{k}{2} \pm \sqrt{k\left(1+\frac{k}{4}\right)},
\end{equation}
so if $k \in [-4, 0]$, then $\left|\ell_+\right| = \left|\ell_-\right| = 1$, and the distance between the trajectories oscillates within some bounds. In the opposite case, when $k \notin [-4, 0]$, for positive $k$, $\left|\ell_+\right| > 1$ and $\left|\ell_-\right| < 1$, so we get $|u_n| \xrightarrow[n\to+\infty]{}\infty$, $|v_n| \xrightarrow[n\to+\infty]{}0$ (and vice versa for negative $k < -4$ or $n\to-\infty$). In general, the eigenvalue $\ell_>: |\ell_>| > 1$ is given by:
\begin{equation}
\ell_> = 1 + \frac{k}{2} + {\rm sgn}(k)\sqrt{k\left(1+\frac{k}{4}\right)}.
\end{equation}
For $k > 0$, the distance between trajectories:
\begin{equation}
d_n = \left|{\boldsymbol d_n}\right| \hspace{-3pt}\underset{n\to+\infty}{\approx}\hspace{-3pt} u_0|\ell_+|^n = u_0e^{n\ln|\ell_+|} = u_0e^{\lambda(0,0) n},
\end{equation}
where $\lambda(0,0) = \ln|\ell_+|$ is the positive Lyapunov exponent at $x,p \hspace{-7pt}\underset{\hspace{-4pt}\mod 2\pi}{=}\hspace{-7pt} 0$ (for $k<-4$, the positive Lyapunov exponent is $\lambda(\pi,0) = \ln|\ell_-|$ and in general, it is given by $\ln|\ell_>|$).

\begin{figure}[t]
	\includegraphics[width=\linewidth ]{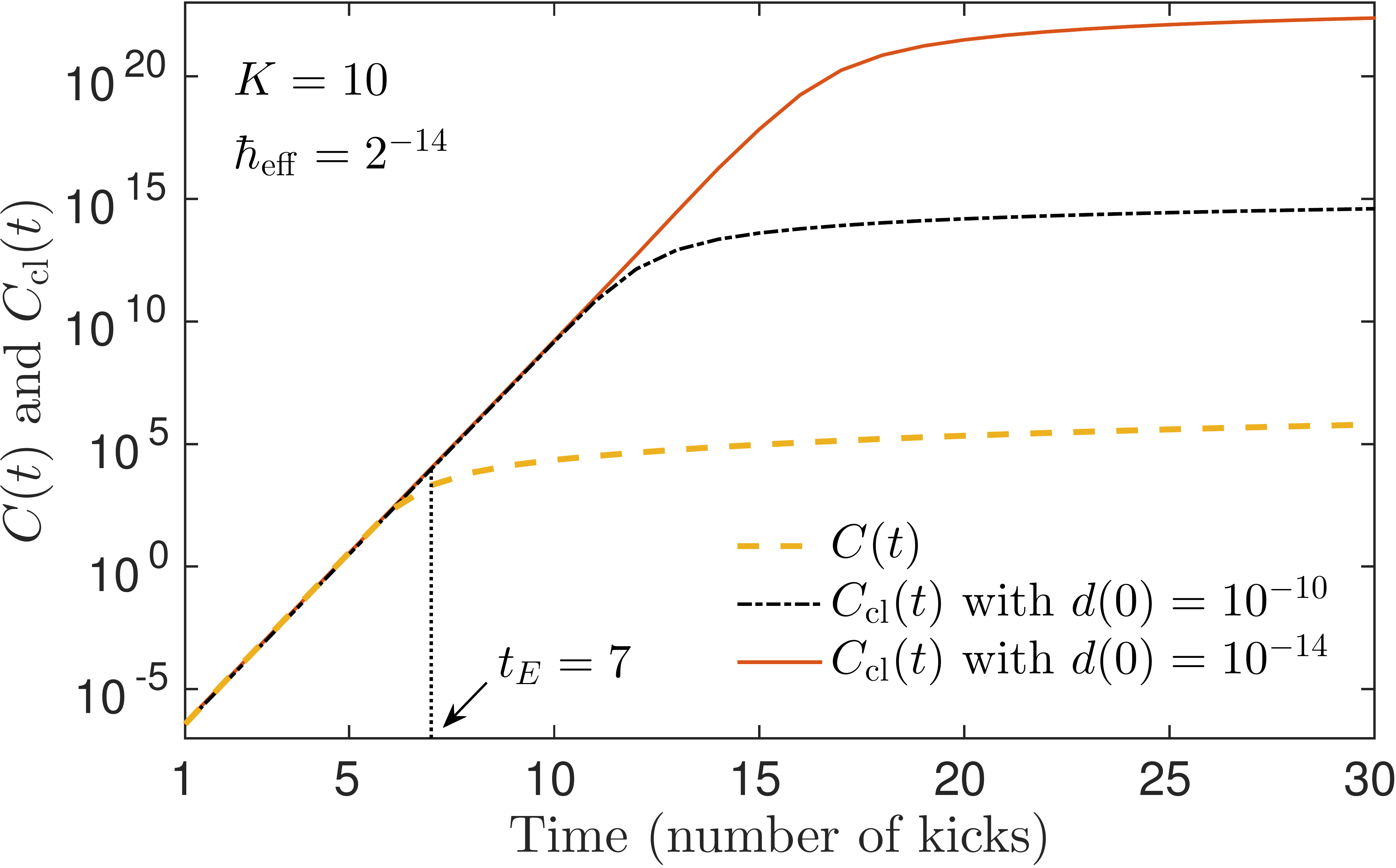}
	\caption{(Color online) $C(t)$ and $\Ccl(t)$ in the semilog scale. The exponential growth of $\Ccl(t)$ saturates due to finite initial distance between trajectories as is shown by comparing $\Ccl(t)$ at $d(0) = 10^{-10}$ and $d(0) = 10^{-14}$. The exponential growth of $C(t)$, however, saturates due to the quantum interference effects that kick in at $t_E = 7$.}
	\label{fig:C_vs_Ccl_K=10} 
\end{figure}
Recall that map (\ref{eq:standard_map_relative_tangent_linear}) only applies to the vicinities of the special points where $k = \pm K$. Let us now average this expression over the whole phase space substituting the general expression $k(x) = K\cos x$ and using $L(x) = |\ell_>[k(x)]|$. Then we arrive to:
\begin{equation}\label{eq:LE_analytical_formula}
\lambda \approx \llangle\lim\limits_{t_c \to \infty}\dfrac{1}{2t_c}\sum\limits_{n=1}^{t_c}\ln \dfrac{u_0^2\ell_+^{2n}+v_0^2\ell_-^{2n}}{u_0^2\ell_+^{2n-2}+v_0^2\ell_-^{2n-2}}\rrangle,
\end{equation}
which upon neglecting the vanishing negative-exponent terms turns into
\begin{equation}
\lambda \approx \llangle\hspace{3pt} \ln \left|\ell_>[k(x)]\right| \hspace{3pt}\rrangle = \llangle\hspace{3pt} \ln L(x) \hspace{3pt}\rrangle,
\end{equation}
that is given explicitly in Eqs. (5, 6). At large $K$, everywhere except the vicinities of $\cos x = 0$, one has: $L(x) \approx |k(x)| = |K\cos x|$, which results in $\lambda \approx \ln(K/2)$.

We point out that the Chirikov's analytical derivation does not yield consistent results for the classical CGR.

{\it Numerical calculation of Lyapunov exponent. ---} The definition can be used directly to calculate LE. However, once chaotic islands become small, the finite initial separation between trajectories (due to numerical limitations) prevents from correct account for the contribution of chaotic trajectories. The tangent map introduced above allows to overcome this difficulty, because it is the derivative of the standard map. The calculation consists in propagating both standard map (\ref{eq:standard_map}) and tangent map (\ref{eq:standard_map_relative_tangent}) using the values of $x_n$ from standard map as inputs for tangent map. Starting at each initial point within some grid, we compute the expression:
\begin{equation} \label{eq:lambda_comp_expr}
\lambda(x_0,p_0) = \dfrac{1}{N}\sum\limits_{t=1}^N\ln\dfrac{d(t)}{d(t-1)}
\end{equation}
for a sufficiently large $N$ and then average it over the phase space. Each time when ${\boldsymbol d}(t)$ becomes large due to exponential stretching, we normalize it to unit length preserving its direction.

\begin{figure}[t]
	\includegraphics[width=\linewidth ]{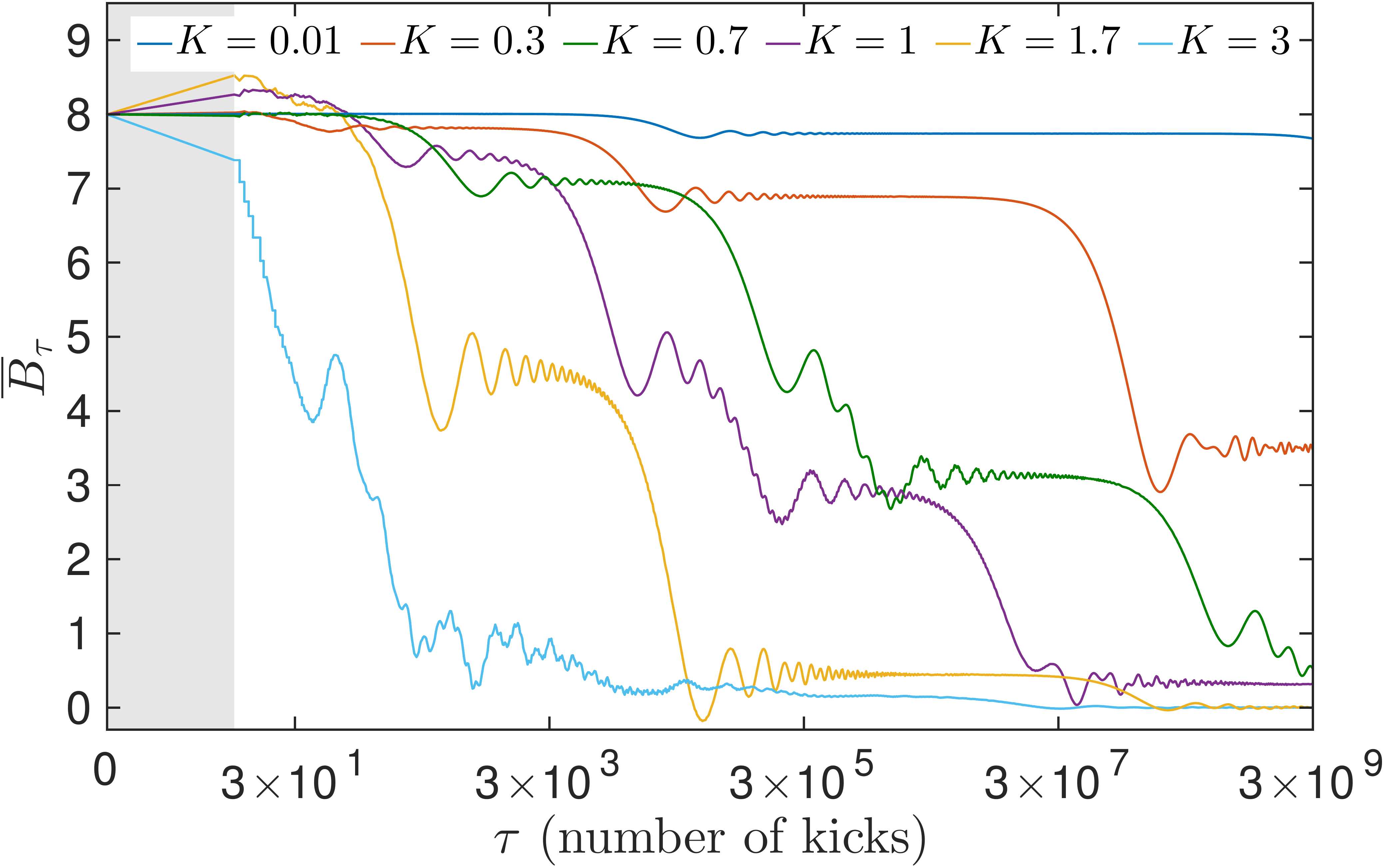}
	\caption{(Color online) $\overline{B}_\tau$ as a function of $\tau$ in the lin-log scale at $K = 0.01$, $0.3$, $0.7$, $1$, $1.7$, and $3$ (from the top to the bottom curve, respectively). The point $\tau = 0$ and the shaded interval $\tau \in (0; 10)$ are added manually to show the initial value corresponding to the wave-packet (4) with $p_0 = 0$ and $\sigma = 4$. At $K = 1.7$ and $3$, one can see the complete relaxation to zero within the range of this plot.}
	\label{fig:Btau_of_tau} 
\end{figure}
{\it Classical CGR. ---} This rate is extracted from the classical analogue of the quantum OTOC:
\begin{equation}\label{eq:Ccl}
\Ccl(t) = \heff^2\llangle\left(\dfrac{\Delta p(t)}{\Delta x(0)}\right)^2 \rrangle \sim e^{2\tilde{\lambda}t}.
\end{equation}
Then
\begin{equation}\label{eq:CGR_calculation1}
\tilde{\lambda} = \dfrac{1}{2}\ln\dfrac{\Ccl(t)}{\Ccl(t-1)}
\end{equation}
Let us average it over some interval in time to improve our fitting accuracy (this step is not necessary though):
\begin{equation}\label{eq:CGR_calculation2}
\tilde{\lambda} = \dfrac{1}{2(t_c-1)}\sum\limits_{t=2}^{t_c}\ln\dfrac{\Ccl(t)}{\Ccl(t-1)}.
\end{equation}
Substituting here $\Ccl(t)$ from Eq.~(\ref{eq:Ccl}) and taking into account that $\Delta x(0)$ is constant throughout the phase space, we obtain:
\begin{equation}\label{eq:CGR_calculation3}
\tilde{\lambda} = \dfrac{1}{2(t_c-1)}\sum\limits_{t=2}^{t_c}\ln\left[\dfrac{\left<\hspace{-4.6pt}\left<\left[\Delta p(t)\right]^2 \right>\hspace{-4.6pt}\right>}{\left<\hspace{-4.6pt}\left<\left[\Delta p(t-1)\right]^2 \right>\hspace{-4.6pt}\right>}\right],
\end{equation}
which is calculated in the same way as the expression (\ref{eq:lambda_comp_expr}) using the tangent map.
The limitation $t \leq t_c$ comes from the fact that, as opposed to the case for LE -- Eq.~(\ref{eq:lambda_comp_expr}) -- rescaling of ${\boldsymbol d}(t) = \dbinom{\Delta p(t)}{\Delta x(t)}$ alters the expressions for CGR -- Eq.~(\ref{eq:CGR_calculation3}) -- and thus is not applicable.

{\it Classical $\Ccl(t)$ vs quantum $C(t)$. ---} In this section, we demonstrate how these corresponding functions compare at large $K$. Fig.~\ref{fig:C_vs_Ccl_K=10} shows the comparison at $K = 10$ (in logarithmic scale). Both $C(t)$ and $\Ccl(t)$ grow exponentially at early times and both slow down after some time. However, in case of $\Ccl(t)$ the reason is purely numerical -- it is calculated by definition here, and the initial distance between the trial trajectories is finite. When the initial separation goes to zero, the exponential growth of $\Ccl(t)$ becomes infinite. On the contrary, the termination of the exponential growth of $C(t)$ is physical and occurs at $t_E = 7$, when quantum interference effects kick in.

{\it Time-averaged two-point correlator as a function of averaging window size. ---} Here, we show the dependence of the time-averaged two-point correlator $\overline{B}_\tau$ on the size $\tau$ of the averaging window at various $K$. In Fig.~\ref{fig:Btau_of_tau}, one can see that the average correlations decay with time in steps, and the larger the kicking strength $K$ is, the faster the correlations decay. Notice that the lin-log scale of the plot implies that the speed-up of this decay with $K$ is exponential.

\end{document}